# Self-ion implantation and structural relaxation in amorphous silicon

J. M. Gibson[*], Rob Elliman[+], T. Susi[o] and C. Mangler[o]

[*]Department of Mechanical Engineering, FAMU-FSU College of Engineering, 2525 Pottsdamer Street, Tallahassee FL 32310, USA (corresponding author e-mail jmgibson@eng.famu.fsu.edu)

[+]Research School of Physics, Australian National University, Canberra, ACT 2601, Australia

[o]Faculty of Physics, University of Vienna, Boltzmanngasse 5, 1090 Vienna, Austria

## Abstract

Self-ion implantation amorphization is an established approach to study the structure and properties of amorphous silicon (a-Si). Fluctuation Electron Microscopy (FEM) has consistently observed Medium-Range Order (MRO) in this system that is not consistent with the Continuous Random Network (CRN) model. Using this technique we find that the degree of MRO first increases on thermal annealing and then decreases before finally recrystallizing. We discuss this new result in the light of previous experimental studies and recent theoretical observations on the favorability of the paracrystalline (PC) model over the CRN in a-Si. At ion doses far above the minimum required to amorphize, a high defect density is found in the PC phase, which anneals out at 500°C. The PC structure after 500°C annealing is independent of the initial implantation conditions and appears to represent a metastable and highly-ordered structure. Higher-temperature annealing causes a reduction in the degree of MRO and the structure approaches but does not reach a fully continuous random network before eventually crystallizing above 600°C. The effect of high dose implantation is to increase the defect density in the as-implanted state and the annealing of these defects is likely responsible for the large characteristic heat evolution at low temperature.

## Introduction

Fluctuation Electron Microscopy (FEM) has been recognized as a powerful tool for detecting medium-range order in amorphous materials [1]. Amorphous silicon (a-Si) has been of particular interest because of its simplicity - as an elemental tetrahedrally-coordinated material - and its importance in semiconductor technology. Amorphous silicon does not form as a bulk material since it is not a glass and instead has been fabricated only in thin-film form. Self-ion implanted silicon is believed to be the purest form of amorphous silicon and correspondingly has been studied extensively. The existence of medium-range order in ion-implanted a-Si has been reproducibly established by Fluctuation Electron Microscopy (FEM) studies[2–6]. However, FEM findings appear to differ on the structure after annealing (relaxation). A recent theoretical study[7] finds evidence that a composite of random network and paracrystallites is of lower energy than a continuous random network in rapidly quenched models of a-Si, consistent with FEM experiments. The question of what happens on annealing of the paracrystallites is of renewed interest and greater significance in the light of this hypothesis.

For many years, based largely on the measurement of Pair Distribution Functions (PDF) from diffraction[8], amorphous silicon was presumed to be a Continuous Random Network (CRN). However simulations have shown that the PDF is unable to clearly differentiate paracrystalline and CRN structures[5]. FEM, through its dependence on higher-order atomic correlation functions, is very sensitive to the medium-range order of paracrystalline structures, or the absence of medium-range order in a continuous random network[9]. FEM studies show that the as-implanted state of a-Si[2-4], in common with other forms of a-Si[10], exhibits a significant degree of paracrystallinity. Earlier work had also shown that annealing of ion-implanted Si can lead to a more disordered, CRN-like structure[2–4]. In contrast more recent work shows that annealing increases medium-range-order[6]. Using a fresh set of self-ion implanted/annealed Si samples, in this work we address the apparent inconsistency and reveal the systematics of the complex evolution of structure after self-ion implantation and annealing. We report for the first time that medium-range order increases on initial annealing up to about 500°C but then decreases in the same samples upon higher temperature annealing. The a-Si annealed at 500°C has the highest degree of MRO - which is the same for different implantation conditions - suggesting it is a metastable state. It seems that as-implanted samples contain medium-range order but with defects that are reduced on annealing. We note the highest defect densities in high-dose implanted samples, far above the doses required for amorphization. Such conditions have been used in calorimetric studies[11] that observed significant heat release on annealing. The annealing of these defects is consistent with the previous explanation of heat release due to defect annihilation[12].

## Experimental

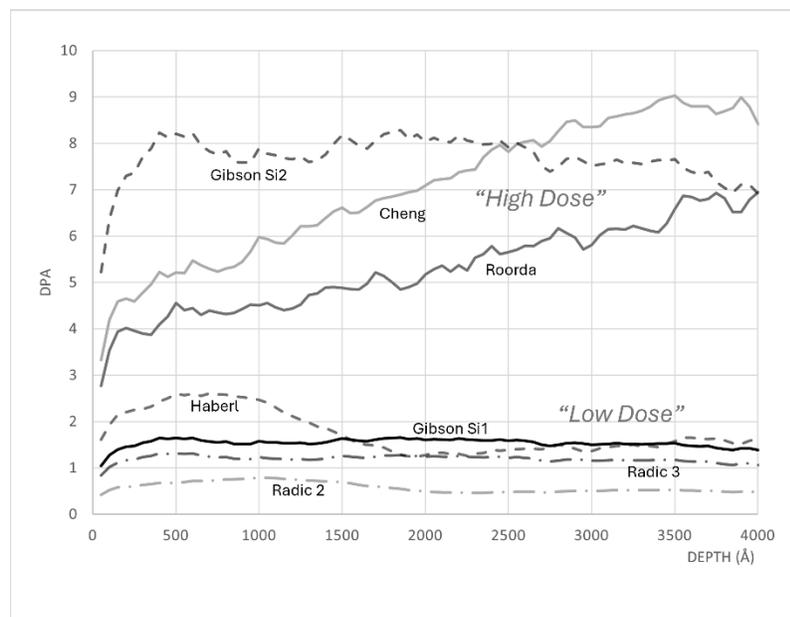

*Figure 1. Depth profiles of damage for self-ion implanted Si samples discussed in this work based on SRIM-2008 calculations (see text).*

Several authors have reported on the structure of amorphous silicon created by self-ion implantation. To make comparisons, it is important to understand the ion effects, characterized primarily by the "displacements per atom" (DPA) as a function of depth, which vary in the studies. It is usually found that approximately 1 DPA is required to amorphize silicon (ion energy could also influence the amorphization process). Figure 1 shows the predicted number of displacements per atom for the silicon amorphization experiments reported in this study, and for several previous publications on the subject. Our calculations used the SRIM-2008 simulation program[13] for modeling the accumulated damage from the ion species, their energies and fluences. All the implantations were carried out with Si ions off-axis and at liquid nitrogen temperatures.

In the following we will discuss several other published studies of medium-range order in ion-implanted Si. For convenience these are referred to in Figure 1 by the first author as follows: Roorda[12], Cheng [2], Haberl [4], Radic2 (double implant) and Radic3 (triple implant) [6].

*Table I. Sample implantation and annealing conditions.*

| Sample | Implant 1 Energy (keV) | Implant 1 Dose (cm$^{-2}$) | Implant 2 Energy (keV) | Implant 2 Dose (cm$^{-2}$) | Implant 3 Energy (keV) | Implant 3 Dose (cm$^{-2}$) | Anneal Temp (°C) | Anneal Time (mins) |
|---|---|---|---|---|---|---|---|---|
| Si1 | 300 | 1.3 x 10$^{15}$ | 150 | 3.6 x 10$^{14}$ | 50 | 2.5 x 10$^{14}$ | None | None |
| Si1a500 | " | " | " | " | " | " | 500 | 60 |
| Si1a560 | " | " | " | " | " | " | 560 | 30 |
| Si2 | " | 6.5 x 10$^{15}$ | " | 1.8 x 10$^{15}$ | " | 1.3 x 10$^{15}$ | None | None |
| Si2a500 | " | " | " | " | " | " | 500 | 60 |
| Si2a560 | " | " | " | " | " | " | 560 | 30 |

Table I shows the implantation and annealing conditions for the samples reported here which were prepared at the Australian National University ion implantation Laboratory (iiLab). The self-ion implantations were conducted at liquid nitrogen temperatures into {100} Si wafers at 7° incidence to the normal. We further note that the Si ions were produced by a SNICS negative-ion source (Source of Negative Ions by Cesium Sputtering), which minimized contamination from isobars such as CO and $N_2$ molecules. For comparison, commercially available sputtered a-Si films (of thicknesses 50, 90 and 150 Å) from SPI™ were also examined in our study and used as thickness standards.

Samples for Scanning Transmission Electron Microscopy (STEM) were prepared by backside chemical thinning, and the areas of study were within a depth of 1000 Å from the original wafer surface (the implanted layer thickness exceeds 0.5 µm). All annealing treatments were carried out before specimen thinning. FEM nanodiffraction patterns were taken at 80kV on a spherical aberration-corrected JEOL™ JEM-ARM200F STEM equipped with a cold field-emission gun. Using the aberration-corrector transfer lens in addition to a range of fixed condenser apertures of 5 – 100 µm in diameter, coherent probes of sizes from 5 Å to 65 Å were produced. Nano-diffraction patterns were recorded with a Gatan ™ Orius CMOS camera with exposure times between 0.1 and 2 s, chosen to keep the electron fluence constant between series. To provide adequate statistics a series of 100 nano-diffraction patterns were taken with each probe size, as the beam was scanned over the sample. We measured the local sample thickness using the amplitude of the elastically scattered signal as a metric, and the known thickness of the sputtered films as a reference. For all samples several series were taken, often at different thicknesses, but the results shown were selected for a uniform thickness of 200 ± 10 Å. Correction for thickness variations[14] used by several authors in the past was found to be unreliable for a-Si and this may explain some differences in the previously reported data. By fixing on a single consistent thickness we remove that ambiguity in making comparisons between individual samples.

From a set of N independent nanodiffraction patterns $I_i(\vec{q}, R)$ recorded as a function of reciprocal lattice vector $\vec{q}$ (q=1/d) and with probe size R, the normalized variance is defined in the usual way[15]

$$V(q,R) = \frac{\sum_{i=1}^{N} I_i^2(\vec{q},R)/N}{\left(\sum_{i=1}^{N} I_i(\vec{q},R)/N\right)^2} - 1 \qquad (1)$$

where the average is over N nanodiffraction patterns at a given probe size (following mode #4 as described by Daulton[16]). The variance data is plotted versus reciprocal lattice vector magnitude q, for several values of the probe size R, and reveals the magnitude, nature and correlation length of any medium-range order. Note that for fully coherent illumination of a truly random sample of atoms, V = 1 with no peaks, and simulations show that a random network also displays negligible structure in V(q,R).

Among the other useful statistical measurements is the electron correlograph based on the normalized autocorrelation function along the azimuthal $\varphi$ axis of a polar plot[17]. The autocorrelation function for one nanodiffraction pattern is

$$G(R,q,\varphi)_i = \frac{<I_i(R,q,\varphi)I_i(R,q,\varphi+\Delta)>_\Delta}{<<I_i(R,q,\varphi)I_i(R,q,\varphi+\Delta)>_\Delta>_\emptyset} - 1 \qquad (2)$$

where $<>_\Delta$ represents integration over the full 360° azimuth for an angular offset $\varphi$. The correlograph

$$C(R,q,\varphi) = \frac{1}{N}\sum_{i=1}^{N} G(R,q,\varphi)_i \qquad (3)$$

is obtained by averaging all images in a series at the same probe size R. The correlograph can reveal the symmetries that are predominant in projections of the medium-range order domains, providing information on their crystalline structure[17].

## FEM Experimental Results

An example of the FEM and normalized variance data versus probe size from one of our samples are given in Figures 2 and 3. The sample is the low-dose Si sample annealed at 500 °C for one hour in $N_2$ (Si1a500). The maximum variance at the first and second peak positions (Figure 3) occurs for a probe size of 27 Å, approximating the medium-range order correlation length[18]. The degree of ordering is related to the peak height over the background[19]. Data was also corrected for the camera modulation transfer function[20] and Poisson shot noise[21]. Figure 2 shows examples of the raw data that goes in to the processing to obtain Figure 3. Fig 2 a) is for a 27 Å probe and Figure 2 b) is for a 9 Å probe. For each series 1) and 2) show two different examples of the one hundred nanodiffraction patterns from the series; 3) is the average intensity of the entire series on a logarithmic scale; 4) is the normalized variance of the entire series from which the graphs in Figure 3 are obtained by annular averaging.

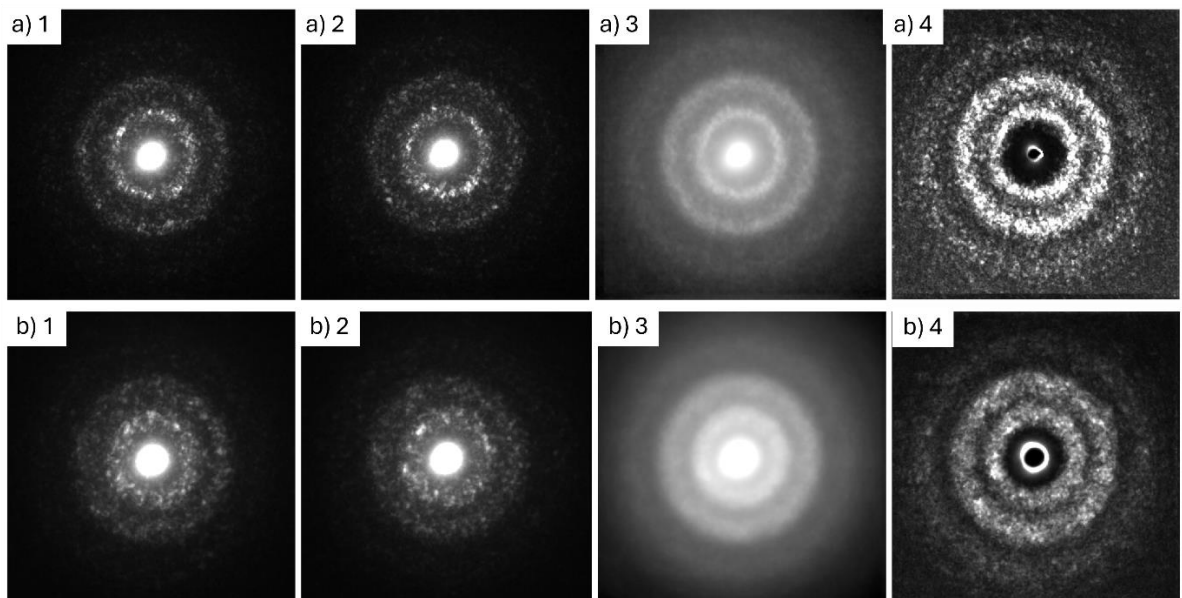

Figure 2: Examples of the experimental data that goes in to producing FIgure 3, for a 27 Å probe (a) and a 9 Å probe (b). In each case 1 and 2 are representative nano-diffraction patterns, 3 is the logarithm of the average intensity of the series and 4) is the normalized variance of all the patterns in the series. Figure 3 shows annular average scans of images such as a)4 and b)4.

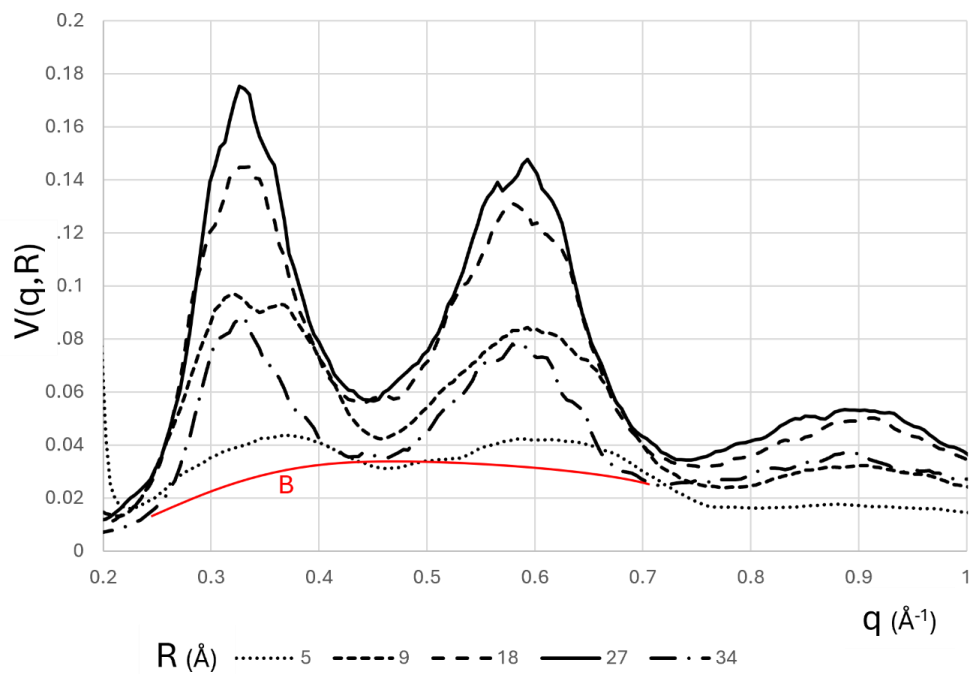

Figure 3 Normalized variance versus probe size R in Å for low-dose annealed sample Si1a500 (the red line B used for background subtraction is referred to in the text. It is extrapolated using a smooth polynomial fit from the dips at 0.23, 0.45 and 0.7 Å$^{-1}$).

The first two strong peaks lie close to 0.31 Å$^{-1}$ and 0.57 Å$^{-1}$, corresponding to the diffuse ring radii seen in diffraction from a-Si. It is further most useful to display the peak heights versus the probe size as a form of "fluctuation map"[1]. Peak heights were obtained by subtracting the fitted background such as shown in Figure 3 for the 34 Å probe (B). Previously published data typically did not employ background subtraction, so the peak heights were larger as a result, but the important conclusions here involve the trends with implantation dose and annealing, which are reflected in both measurements. (We have found that background-subtracted peak heights are more resilient to changes in the noise subtraction and exposure conditions than the peak height itself, which we will discuss in more detail in a future publication.) Fluctuation maps, such as that obtained from the data in Figure 3, shown in Figure 4, directly reveal the correlation length (from the peak in the variance where the probe size matches the correlation length [17,18] and the degree of ordering which is related to the peak height[1].

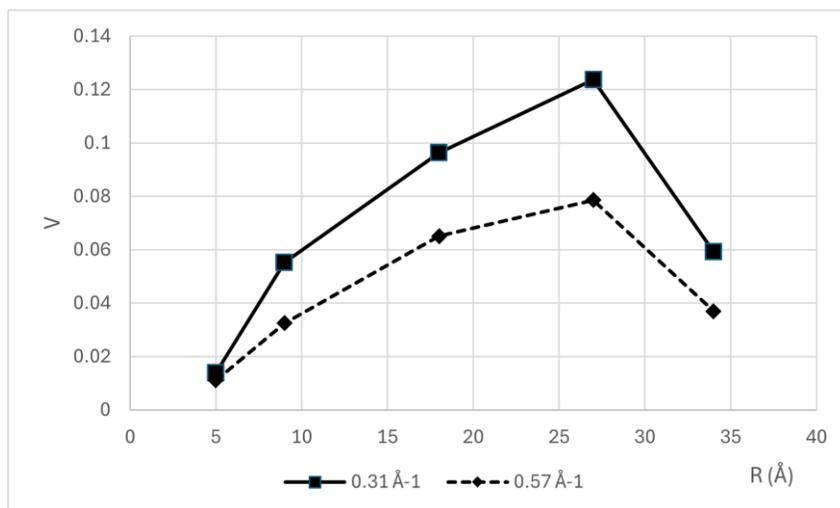

*Figure 4. A fluctuation map showing the normalized variance V versus q and probe size R at the two values of q representing the strongest peaks after background subtraction.*

Data for all the samples listed in Table I is presented in Figure 5. The two prominent normalized variance peak heights at approximately 0.31 Å$^{-1}$ (first peak) and 0.57 Å$^{-1}$ (second peak) are shown for the 27 Å probe size for all samples. Typically, several data series were recorded per sample, giving the error bars shown on the plots in Figure 5. Table II gives quantitative data from samples that will be discussed in the text.

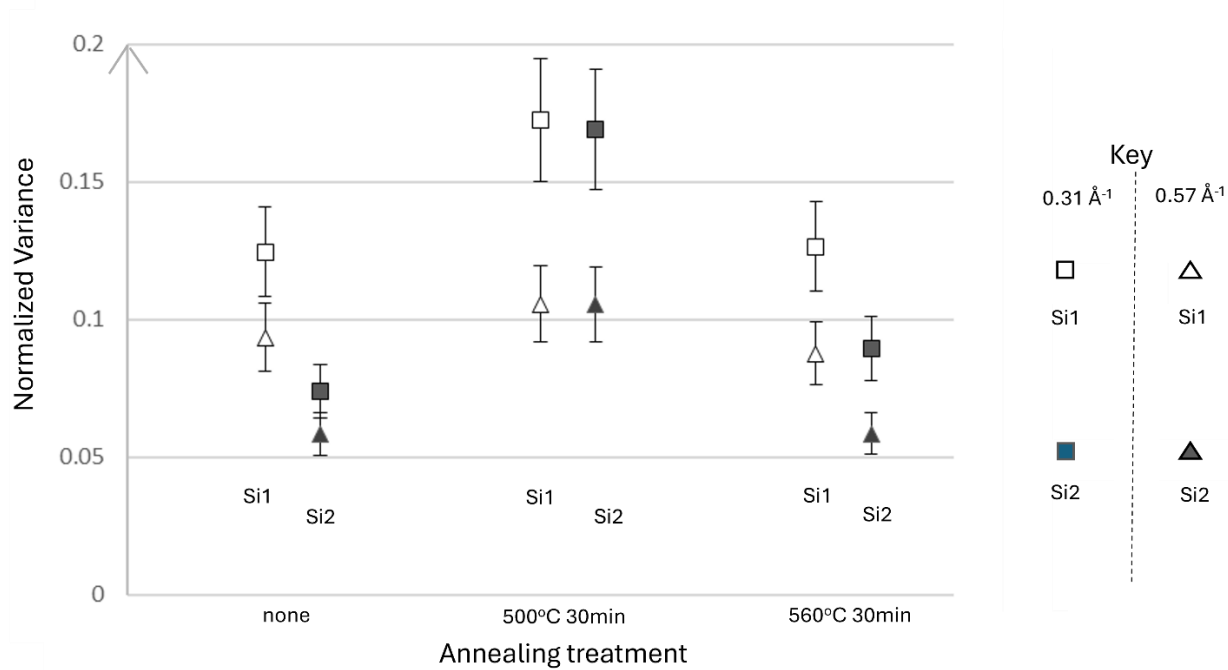

*Figure 5 A summary of the peak heights for all the Si samples in Table 1 for a probe size of 27Å at a thickness of 200 Å.*

| Sample | 0.31 Å$^{-1}$ Peak 1 height | 0.57 Å$^{-1}$ Peak 2 height | Peak Height Ratio 1/2 | $\Lambda_1$ (Å) Correlation Length Pk 1 | $\Lambda_2$ (Å) Correlation Length Pk 2 |
|---|---|---|---|---|---|
| Si1 | 0.12 *(.02)* | 0.094 *(.01)* | 1.9 *(.1)* | 21.6 *(.05)* | 21.4 *(.05)* |
| Si1a500 | 0.17 *(.02)* | 0.11 *(.01)* | 2.3 *(.1)* | 22.0 *(.05)* | 21.9 *(.05)* |
| Si1a560 | 0.13 *(.02)* | 0.087 *(.01)* | 2.0 *(.1)* | 21.8 *(.05)* | 21.5 *(.05)* |
| Si2 | 0.074 *(.01)* | 0.058 *(.005)* | 1.8 *(.1)* | 18.7 *(.05)* | 19.1 *(.05)* |
| Si2a500 | 0.17 *(.02)* | 0.11 *(.01)* | 2.2 *(.1)* | 21.1 *(.05)* | 20.4 *(.05)* |
| Si2a560 | 0.09 *(.01)* | 0.058 *(.005)* | 1.5 *(.1)* | 22.6 *(.05)* | 21.8 *(.05)* |
| Sputtered a-Si | 0.091 *(.01)* | 0.077 *(.01)* | 1.7 *(.1)* | 20.4 *(.05)* | 20.3 *(.05)* |

*Table II. Normalized variance peak heights (for a 27 Å probe) and correlation lengths for the samples examined in this study. Peak heights are background subtracted.*

The correlation lengths (Λ) in Table II are obtained through the weighted average $\Lambda = \frac{\sum R_i V_i}{\sum V_i}$ of the probe sizes $R_i$ and the peak heights $V_i$. The data reveals that the two samples annealed at 500°C have the greatest degree of medium-range order. Further clarification of the nature of the medium-range order is provided from the experimental correlographs. Figure 6 shows an experimental correlograph as a function of azimuthal angle integrated over the 0.31-0.34 Å$^{-1}$ range from the Si1 sample. There are prominent peaks near 70.5° and equivalent angles between {111} planes in the <110> projection of diamond Si. Peaks at this approximate position were observed in all the samples. For the 0.57 Å$^{-1}$ normalized variance peak correlographs were noisier but revealed azimuthal correlation peaks near 60° and 120°, characteristic of interplanar angles between 220 reflections.

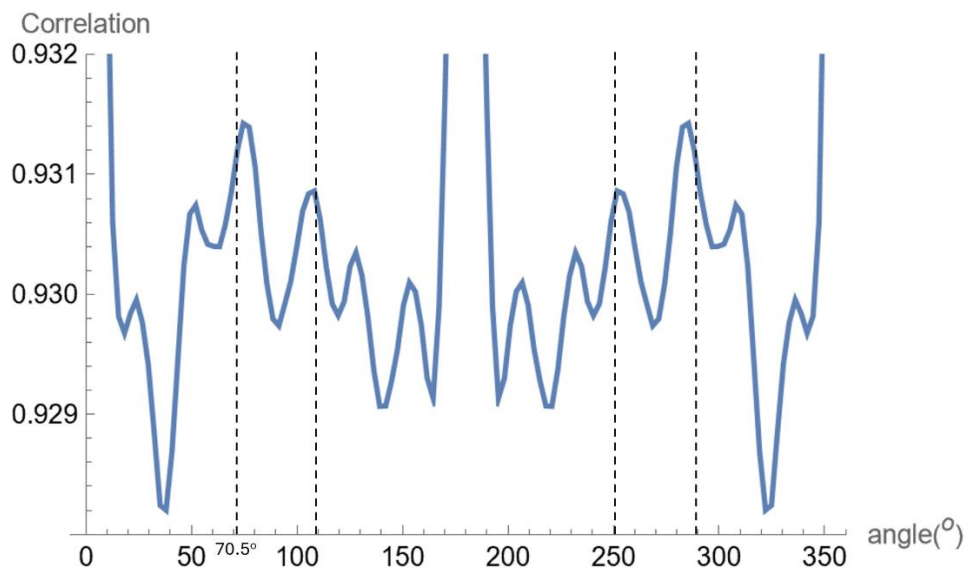

*Figure 6. Angular correlations near the 111 position reveals the presence of ordered paracrystallites in the Si1 sample. The strongest peak, except for the "Freidel" peak at 180°, is near to the tetrahedral angle 70.5° between {111} planes in the <110> projection. (The vertical dotted lines also show symmetry related tetrahedral angles).*

## Discussion

### As-implanted samples

The first FEM measurements of ion-implanted amorphous silicon were reported by Cheng [2,3]. Comparing experiments with simulations, they showed that the structure of as-implanted Si was better described by the paracrystalline model, with significant medium-range order, and not by a pure continuous random network. All subsequent FEM measurements have observed this result to varying degrees [4,5,6]. Variations in the degree of medium-range order depend largely on the implantation and annealing conditions and it became clear there was a need for further systematic study over a range of annealing parameters and implantation doses to better understand this behavior.

We have examined two sets of samples, one at a relatively low ion dose, just sufficient to amorphize (~1.5 DPA) and the other at a relatively high dose (~8 DPA) far above the threshold. In all of the samples we observe that the FEM peaks at 0.31 Å$^{-1}$ and 0.57 Å$^{-1}$ have a height ratio close to 2:1 and correlation lengths of order ~20Å, with only small variations in those parameters. By contrast peak heights depend more strongly on the implantation and annealing conditions. The correlation length implies a paracrystallite "grain size" in the vicinity of 20 Å, which is also consistent with simulations of the size dependence of the peak height ratio (~2:1) for diamond paracrystallites[14,19]. The correlographs are similar to the diamond structure, although some additional peaks are found.

Because of the relative consistency of correlation length and peak-height ratio, we attribute the peak-height variations to defect density and/or MRO volume-fraction variations. Of the as-implanted samples, the low-dose sample Si1 has higher MRO demonstrated by the normalized variance peak heights. Radic-3 is close to our low-dose condition and their FEM results are similar. (Radic-2 was not fully amorphized and so the results there are more complex, especially on annealing where homogeneous epitaxial recrystallization interferes with the process.)

We attribute a high defect density to the significantly lower variance peak heights we find in the high-dose implanted sample Si2. Note that with FEM we observe the defects through their effect on the sharpness and intensity of variance peaks. Defects may also be present in the random network component of the sample, but these cannot be easily observed either by FEM or by diffraction-based measures such as the PDF. A slight difference in the first peak width of the PDF observed in evaporated films[22] and ion-implanted films[12] after annealing could now be attributed to the paracrystalline component of the composite, which was not considered, rather than to the random network component. The peak height ratio and correlation length in our high-dose sample are also slightly smaller, which would be consistent with slightly smaller PC grains (~10% difference). The implantation conditions for these samples were closer to those used by Haberl[4], Cheng[2] and Roorda[12].

Finally, note that the sputtered a-Si films have similar correlation length and peak heights as the low-dose sample Si1, demonstrating the fundamental similarity of the PC structure in very different a-Si samples.

## Temperature Dependence of MRO

In agreement with Radic[6] we found that annealing the low-dose sample to 500°C increases the degree of MRO. The correlation length is unchanged, but the peak height ratio increases slightly. The correlation length measured by the probe size with the highest variance (Table II) is almost the same as seen by Radic[18] using a similar method. A reduction in the defect density within the paracrystals and possibly an increase in their number density could account for the results. The concentration of paracrystals is difficult to measure because of experimental uncertainties, for example from decoherence[23], but we can put a lower limit on this by noting that a fully coherent simulation with 18 Å diamond paracrystallites produces $0.31 Å^{-1}$ background-subtracted peak heights of 0.15 at a concentration of 5% by volume[9].

Our results show that annealing of both the low-dose Si1 and high-dose Si2 samples leads to very similar structure. The FEM parameters for Si1a500 and Si2a500 are the same within experimental error, even though the as-implanted samples are quite different. This suggests that there is a metastable state of paracrystalline Si that can be achieved after moderate annealing of amorphous Si, independent of the initial conditions. Such an observation based on the stability of the properties of annealed amorphous silicon was made long ago by Tsu[24]. While this "anneal-stable" form of a-Si was thought to be a perfect CRN, our results, together with recent theoretical considerations[7] suggest instead that the structure comprises well-ordered paracrystallites in a random network matrix.

Our observations agree with all previous FEM studies on the effect of higher temperature annealing, above 500 °C[2–4]. Table I shows that 560°C annealing reduces the degree of medium-range order compared to the 500 °C annealed samples. We are limited from studying temperatures approaching 600°C by competition with epitaxial regrowth from the substrate, which is exacerbated by the highest energy implant we used of only 300 keV (other studies employed implants of up to 1 MeV to create thicker layers). However, it seems likely that the degree of MRO would continue to decrease as the annealing continues at higher temperatures, based on earlier findings[3]. The correlation length does not decrease on higher temperature annealing, although there is a slight reduction in the peak-height ratio. This would suggest the possible dissolution of some but not all of the paracrystals under these conditions. Based on our observations we conclude that the critical nucleus size for homogeneous recrystallization at 580°C must exceed 20 Å, since the the paracrystals are not growing.

While our data agrees with prior literature that MRO decays after high-temperature annealing, there is published evidence of such a decay at 500°C and below, which is inconsistent with our recent findings. It could be that the exact profile of ion damage plays some role. For example our experiments and those of Radic[6] employed lower ion energies (50keV) to amorphize the surface regions that were studied, whereas those who saw low temperature MRO reduction on annealing used higher doses of more energetic ions, for example 80keV [4] and 300keV[2] to amorphize the surface. The latter study also reported significant depth dependence of the MRO, indicating the possible complexity of a multiple implant system. Errors such as inadequate thickness or Poisson noise compensation may have also affected earlier works. Nevertheless, our current studies here replicate many other aspects of previous work and, to our knowledge, these are the first experiments to include ion dose as a variable, revealing in particular the substantial effect of high-doses on defect density. This is particularly important because all of the calorimetric studies so far have used high doses to observe the heat release from amorphous silicon during relaxation[11] and before crystallization. The calorimetric studies require large volumes and were carried out by necessity under high dose conditions similar to the high-dose sample Si2 (for example in the key paper of Roorda[12]), and required multiple implant energies.

The paracrystalline structure comprises a set of imperfect and strained nanocrystals (paracrystallites) embedded in a random network matrix, as seen in the recent theoretical models of Rosset[7]. A diamond crystal of 20 Å in size would contain about 400 atoms. At a density of 5% these would be spaced on average by about 40 Å. The random network in which they are embedded is not continuous- in fact, the structure would be strongly influenced by the strain field from the paracrystals and theory[7] suggests they could lower the energy of a random network by their presence. It would be interesting to examine structures such as produced by Rosset[7] to probe their simulated FEM properties, such as correlation length and peak height ratios.

Theoretical work is pivotal in recognizing the relative stability of the PC structure compared with the CRN[7]. However the authors of that study recognized that their structures represent the extremely rapid quenching rates typical of "laser-glazed" a-Si and not of the samples discussed in our paper. Under rapid quenching conditions they found the typical size of paracrystallites to be ~40 atoms or less, equivalent to a correlation length of ≤ 12 Å. In a possibly similar situation, pressure-induced a-Si has been produced that has substantially less MRO than implanted and annealed a-Si. On thermal annealing the MRO increases to match that of annealed implanted Si[4]. Therefore the ~20 Å correlation length that we and others[18] observe which is equivalent to ~200 atoms could be more representative of the metastable state of the amorphous silicon structure.

Ultimately, it would be of great interest to visualize real atomic positions in a silicon sample using 3D TEM atom tomography which has very recently become feasible[25]. Since network topology controls defects, even if the paracrystallite density is only 5% it is likely that some physical properties could differ significantly from a putative Continuous Random Network.

## Heat release and relaxation

The substantial heat releases in amorphous silicon have, to our knowledge, only been confirmed in the very highly damaged samples considerably above the threshold for amorphization, and may not be intrinsic to the ion-amorphization process. To obtain the volume to do calorimetric and other measurements, substantially higher than threshold doses were required.

Roorda[12] carried out a thorough analysis of the effect of annealing on ion-implanted Si, at a dose similar to our high-dose sample and the previous studies by Cheng[3]. They observed a significant heat evolution at and below 500°C and were able to fit the kinetics to a bimolecular defect annihilation model similar to crystalline Si damage annealing. Our current results where we observed a significant defect density reduction in the paracrystalline structure are quite consistent with this explanation. Cheng[3] had observed with FEM a constant reduction in MRO on annealing and attributed the heat release to this process. Our recent work would support Roorda's claim that the heat release, at least at temperatures ≤ 500°C, does not originate from disordering but from defect annihilation[26]. We can only detect this phenomenon in the paracrystalline regions but it may also be occurring in the random network.

On higher temperature annealing all FEM studies agree that disordering does occur as the paracrystalline structure loses MRO. Relaxation and heat evolution continues at the highest temperatures before crystallization. When using rapid thermal annealing it has been confirmed that relaxation and heat evolution continues up to 850°C[12]. In this arena it seems that Cheng[3] are correct that structural transformation towards a more Continuous Random Network is the culprit.

On the annealing of amorphous Si, X-ray diffraction has shown a slight improvement in the ordering represented by sharpening of the first peak in the radial distribution function and a narrowing of the bond-angle distribution[22, 12]. Our measurements reproduce this observation but indicate that this improved ordering for annealing at ≲ 500°C is associated with reduction in the defect density of the initial paracrystalline structure and not necessarily from the random network component. Using model simulations we can reproduce similar pair distribution effects seen with X-rays[12,22] simply from improved ordering in and around the paracrystallites without significant changes in the random network component.

It seems that the results of previous work can be rationalized in the context of our data. Ion implantation creates a very defective state, but it is a defective paracrystalline state (a compact of paracrystals and random network) and not a pure CRN. The well-known heat release up to about 500°C in Si is well-explained by the defect annihilation model and leads to a better ordered paracrystalline state. However, on further annealing, the degree of paracrystallinity drops, perhaps as a result of entropy favoring a more disordered state.

# Conclusion

We have studied the effect of ion-implantation dose and annealing conditions on medium-range order in self-ion implanted amorphized silicon. All samples display some degree of paracrystalline MRO, consistent with a composite of ~20 Å sized paracrystallites in a random network matrix. We confirm recent reports that low-ion-dose samples, just above the threshold for amorphization, have increased MRO on annealing at ~500°C[6]. For high-ion-dose samples, annealing creates a very similar structure, but the as-implanted structure is significantly more disordered. The data is consistent with a high defect-density in the high-dose damaged PC structure. These defects are almost certainly those identified as responsible for heat release on annealing[12]. Our results support the conclusion that defect reduction, and not relaxation of the amorphous structure is the origin of low-temperature heat release[3,26].

On annealing at around 500°C the material takes on its most ordered paracrystalline configuration, apparently independent of implantation conditions. This is consistent with recent work that has shown the increase in ordering after such annealing for low-dose implants[6]. Experimental data on angular correlations supports the conclusions that the paracrystals have diamond-like coordination and a correlation length of ~20 Å. It is difficult to estimate accurately the volume fraction of paracrystals in the composite, but it is a minimum of 5% and could be higher. Even though the density is low, theory suggests that the paracrystalline regions are intrinsic to amorphous Si and may help to stabilize the random network[7]. They could also play a very significant role in the defect structures and material properties.

On higher temperature annealing below the temperature at which crystallization dominates ($\lesssim$ 600°C) the degree and nature of ordering diminishes as previously observed[2,3,4]. We attribute this to a competition with entropy which favors a disordered structure. We hypothesize that if crystallization could be impeded, the structure may move closer to a random network but likely always contains some degree of paracrystallinity. This confirms many previous reports that at higher temperature annealing (~550°C) the degree of MRO is decreased in high ion dose samples, and confirms the same effect in low dose samples under the conditions employed by Radic[6]

High-dose implanted samples (DPA >> 1) exhibit a high degree of disorder, with a defective paracrystalline structure. Annealing these samples brings them to the state described for low-dose implantations – well-ordered paracrystals. This may resolve a previously held disagreement between the FEM measurements and calorimetric kinetics postulated by Roorda [12] These authors elegantly showed that bimolecular kinetics could explain the origin of heat evolved by high dose ion implanted samples, with very similar kinetics to defect annealing in crystalline Si. We believe these defects exist in the paracrystalline state and we see they are removed on annealing. However, the partially annealed state is not entirely a CRN as Roorda supposed, but a better ordered paracrystalline structure embedded in a CRN. It is possible that more heat is evolved at high temperatures by reduction in the ordering, but we have no evidence for this speculation.

There remain many questions about the paracrystalline state and its thermodynamics and kinetics in amorphous Si. It is clearly a more complex and intriguing system than at first believed, offering new parameters for microstructural control, and resulting properties. The exploration of reverse Monte Carlo simulations including angular correlation experimental data, such as the correlograph or related approaches[27], and FEM at multiple probe sizes could provide greater insight into the paracrystalline structure.

## Acknowledgements

The authors gratefully acknowledge support from the FAMU-FSU College of Engineering and the Department of Physics at the University of Vienna. Electron microscopy was performed at the National High Magnetic Field Laboratory, which is supported by National Science Foundation Cooperative Agreement No. DMR-2128556 and the State of Florida. Assistance from Dr. Yan Xin (MagLab), & Kevin McIlwarth and Nathaniel Didier (JEOL Inc) in configuring the JEOL ARM200F was invaluable. Assistance with specimen preparation from Andreas Berger at University of Vienna is much appreciated. Finally, we acknowledge the Australian National University ion-implantation Laboratory (iiLab), a node of the NCRIS Heavy Ion Accelerator Capability, for access to ion-implantation facilities. JMG is particularly grateful to Zhiyong "Richard" Liang, Director of the High Performance Materials Institute for his advice and support.

## Author Declarations

Conflicts of interest: The authors have no conflicts to disclose.

Author Contributions: **J.M. Gibson** – Conceptualization (lead), Visualization, Data Curation, Formal Analysis, Writing/Original Draft Preparation; **Rob Elliman** – Conceptualization (supporting), Resources, Writing/Review and Editing; **T. Susi** – Conceptualization (supporting), Writing/Review and Editing; **C. Mangler** – Investigation, Software, Methodology, Writing/Review and Editing

## Data Availability

The data that support the findings of this study are available from the corresponding author upon reasonable request.

## Supplementary Materials:

### Experimental Methods:

Images were analyzed with Mathematica™ scripts, which can be made available to those who are interested by contacting the primary author at *jmgibson@eng.famu.fsu.edu*. The data processing included filtering to ensure a narrow range of thickness (±5-10%), removing any scattering from large nanocrystals or contamination, removing the background shot noise and correcting for the MTF of the camera. For noise removal we calculated the number of counts per electron based on shot noise measurements in an MTF-corrected uniformly-illuminated field on the camera, following the algorithm from Fan[20]. The reference SPI a-Si films are very clean, showed very little contamination, and exhibited less than 10% standard deviation in thickness. We used the absolute amount of scattering into the first-ring of the diffraction patterns as a proxy for thickness – in samples much less than one mean-free-path in thickness this is linearly proportional to the thickness. Our probe sizes are based on the Rayleigh criterion using the semi-angle of convergence.

One significant processing difference from our previous reports is that we display the peak height over the background level, instead of the absolute peak height. Several authors have recognized that this is a more robust measurement, since the background level changes, with for example exposure, but the peak to background ratio is less affected. We will discuss this in more detail in a separate paper, where we attribute these differences to decoherence effects associated with specimen motion during exposure, and to the point-spread function of the camera. Simulations confirm that the peak height to background is a robust measurement, less affected by the camera MTF and other experimental parameters.

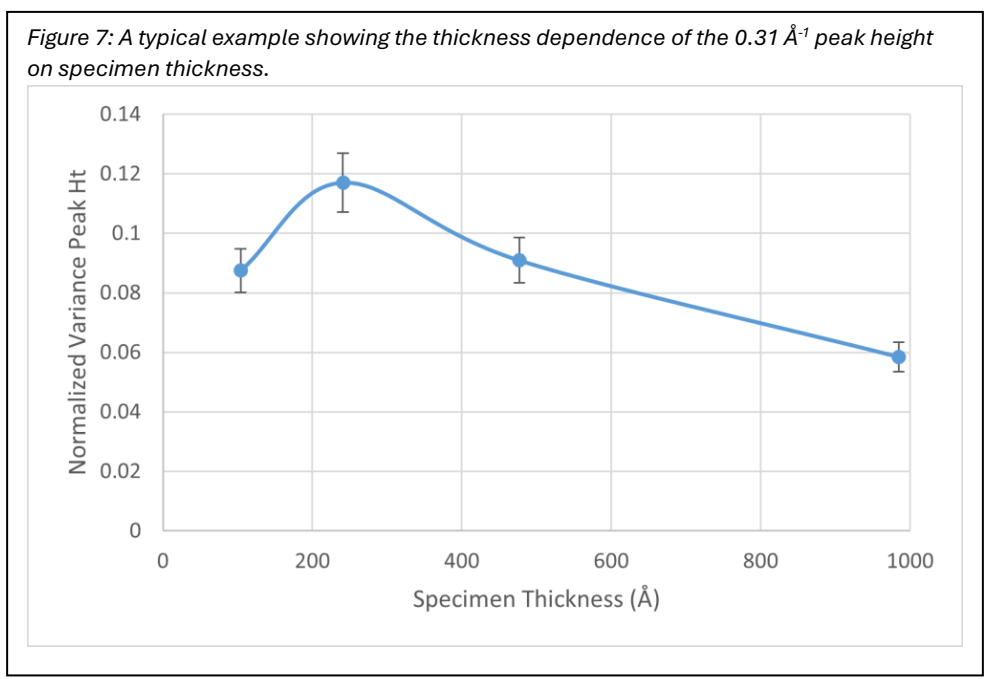

Figure 7: A typical example showing the thickness dependence of the 0.31 Å$^{-1}$ peak height on specimen thickness.

As shown in Figure 7, the thickness dependence of the variance peak heights does not follow the simple ~*1/t* result seen from simulations and commonly used for thickness correction[14], except at higher thicknesses much above 200 Å . As a result, we made several measurements of *V(q,R)* at different thicknesses, and used the data such as in Figure 1 to extrapolate to a constant thickness, which we chose to be 200 Å. This process is the primary origin of the error bars shown in Figure 4.

While all the data published here was taken at 80keV in the JEOL ARM200F, initial work was carried out at 60keV in the Nion UltraSTEM 100[31] located in the Faculty of Physics, Physics of Nanostructured Materials Group at the University of Vienna's Sternwarte Laboratory in Vienna, Austria.